\documentclass[aps,pra,twocolumn,showpacs,superscriptaddress]{revtex4}

\usepackage{graphicx}
\usepackage{amsmath}
\usepackage{amssymb}

\input{epsf}
\begin{document}

\title{Stability of Inhomogeneous Multi-Component Fermi Gases}

\author{D. Blume}
\affiliation{Department of Physics and Astronomy,
Washington State University,
  Pullman, Washington 99164-2814}
\affiliation{
JILA, University of Colorado,
Boulder, CO 80309-0440}
\author{Seth~T. Rittenhouse}
\affiliation{Department of Physics and JILA, University of Colorado,
Boulder, CO 80309-0440}
\author{J. von Stecher}
\affiliation{Department of Physics and JILA, University of Colorado,
Boulder, CO 80309-0440}
\author{Chris H. Greene}
\affiliation{Department of Physics and JILA, University of Colorado,
Boulder, CO 80309-0440}

\date{\today}

\begin{abstract}
Two-component equal-mass Fermi gases, in which unlike atoms interact through
a short-range two-body potential and like atoms do not interact,
are stable even when the interspecies $s$-wave scattering 
length becomes infinitely large. 
Solving the many-body Schr\"odinger equation within a hyperspherical
framework and by Monte Carlo techniques,
this paper investigates how the properties of 
trapped two-component gases change if a third or fourth component
are added.
If all interspecies scattering lengths are equal and negative, 
our calculations suggest that
both three- and four-component Fermi gases become unstable
for a certain critical set of parameters.
The relevant
length scale associated with the collapse is set by the 
interspecies scattering length and we argue that
the collapse is, similar
to the collapse of an attractive trapped Bose gas, a many-body phenomenon.
Furthermore, we consider a three-component Fermi gas in which two
interspecies scattering lengths are negative while the
other interspecies scattering length is zero.
In this case, the stability of the Fermi system 
is predicted to depend
appreciably
on the range of the underlying two-body potential.
We find parameter combinations for which the system 
appears to become
unstable
for a finite negative scattering length and parameter combinations
for which the system
appears to be made up of
weakly-bound trimers that consist of one fermion of each
species. 
\end{abstract}


\maketitle

\section{Introduction}
\label{introduction}
Over the past decade or so, the field of ultracold gases has 
seen tremendous breakthroughs. After
reaching degeneracy in Bose~\cite{ande95,davi95} 
and Fermi~\cite{dema99} gases,
the realization of an atom laser~\cite{mewe97,bloc99,hagl99}, 
of the Mott-insulator 
transition~\cite{grei02},
and of the conversion from an atomic to a molecular gas~\cite{donl02}
followed.
Many of the present-day studies take advantage of the tunability of the
atom-atom scattering length in the vicinity of a so-called
Fano-Feshbach resonance~\cite{stwa76a,ties93}.
As an external magnetic field is tuned through
its resonance value, the sign of the scattering length 
changes~\cite{inou98,corn00}. 
Exactly on resonance, the scattering length
is infinitely large, allowing for the study
of strongly-correlated systems. Experimentally, the speed of the magnetic 
field ramp can be changed, allowing adiabatic ramps, 
for example, 
and the ramp itself can be reversed.
It is this versatility that made possible the experimental study of the BCS-BEC 
crossover, 
using ultracold atomic Fermi gases 
trapped in two different hyperfine states ~\cite{grei03,zwie03}.

Using present-day technology, the realization of degenerate
multi-component
atomic Fermi gases appears possible in principle. 
The occupation of more than two different hyperfine states
of the same species requires, neglecting for the moment possible losses,
only moderate changes of current set-ups.
A particularly promising candidate appears to be $^{6}$Li~\cite{abra97},
and the coexistence of three hyperfine states has already been demonstrated
for $^{40}$K~\cite{rega03b}. 
Alternatively,
a number of goups are presently pursuing the simultaneous trapping of
three different atomic species~\cite{tagl06,grimm}.
In the former scenario, the atomic masses of all components are
equal, whereas in the latter scenario, the atomic masses of the 
different components differ.
In either of these realizations 
of multi-component Fermi gases, all or some of the
interspecies scattering lengths may be tunable thanks to the 
possible existence of
magnetic or optical Fano-Feshbach
resonances. This may open the possibility
to experimentally investigate the stability and to study,
provided an extended stable regime exists, 
the behaviors of multi-component Fermi gases
as a function of the $s$-wave scatterig length.

Using two
different theoretical frameworks,
this paper considers three- and four-component
Fermi gases, 
and compares
their behaviors with those
of two-component Fermi gases.
In particular, we ask how the
stability 
of two-component Fermi gases changes when a third or fourth
component are added.
It is now well established that 
trapped two-component Fermi gases
are stable 
even when the interspecies
$s$-wave scattering length 
is negative and its magnitude
is infinitely 
large~\cite{bake99,heis01,carl03,astr04c,footnotestable}.
The stability of inhomogeneous as well as 
of homogeneous two-component Fermi gases
with 
attractive short-range interactions and arbitrary
interspecies scattering length can be attributed to the
Pauli exclusion principle (also referred to as Fermi pressure), 
which introduces effective repulsive intraspecies interactions
that more than compensate the attractive interspecies
interactions.
In contrast, homogeneous Bose gases with
negative scattering lengths
are unstable;
they
can, however, be stabilized by an external confining potential as long as the
product of the number of bosons and the $s$-wave scattering length
is less negative than a certain critical value~\cite{dodd96,donl01,robe01}.

Section~\ref{sec_system} introduces the Hamiltonian of
trapped multi-component
Fermi gases as well as a simple ``counting argument'' that 
turns out to be quite
useful in understanding the stability of multi-component Fermi
gases.
Section~\ref{sec_hyper} investigates the
stability of three- and four-component equal-mass Fermi gases within a 
hyperspherical framework, focussing on the large and small particle
number limits.
The physical picture emerging from the hyperspherical treatment is 
further investigated
in Sec.~\ref{sec_mc}, which solves the many-body Schr\"odinger
equation for short-range interactions using Monte Carlo techniques.
Finally, Sec.~\ref{sec_conclusion} summarizes and connects our results
with those available in the 
literature~\cite{moda97,heis01,hone04,hone04b,he06,sedr06,zhai07,paan07,cher07,chan06,beda06,ritt07,lech05,guan07,liu07}.

\section{Hamiltonian and general considerations}
\label{sec_system}

The Hamiltonian $H$ for an atomic Fermi gas with $\chi$ components
under external spherically symmetric harmonic confinement is given by
\begin{eqnarray}
\label{eq_ham}
H =
\sum_{\alpha=1}^{\chi}\sum_{i=1}^{N_{\alpha}} \left(
-\frac{\hbar^2}{2m_{\alpha}}
\nabla_{\vec{r}_{\alpha i}}^2 +
\frac{1}{2} m_{\alpha} 
\omega_{\alpha}^2 \vec{r}_{\alpha i}^2 
\right) +
\nonumber \\
+ \sum_{\alpha < \beta}^{\chi} 
\sum_{i=1}^{N_{\alpha}} \sum_{j=1}^{N_{\beta}} 
V_{\alpha \beta}(|\vec{r}_{\alpha i} - \vec{r}_{\beta j} |).
\end{eqnarray}
Here, the number of 
atoms of the $\alpha$th component is denoted by $N_{\alpha}$,
and the total number of atoms is given by $N$, 
$N = \sum_{\alpha=1}^{\chi} N_{\alpha}$.
In Eq.~(\ref{eq_ham}), 
$\vec{r}_{\alpha i}$ denotes the position vector of the $i$th
atom of the $\alpha$th component, 
measured with respect to the center of the trap,
and $m_{\alpha}$ and
$\omega_{\alpha}$ respectively
the atomic mass and the angular frequency
of the $\alpha$th component.
The potential
$V_{\alpha \beta}$ describes the interaction between
an atom of the $\alpha$th and an atom of the $\beta$th component.
This work considers
a 
zero-range potential (see towards the end of this section
and Sec.~\ref{sec_hyper})
and a purely attractive short-range potential 
(see Sec.~\ref{sec_mc}). 
In both cases, we characterize the strengths of the $V_{\alpha \beta}$
by the $s$-wave scattering lengths $a_{\alpha \beta}$.
We assume that the two-body interactions are
independent of spin, implying that the number of atoms in each spin
state is conserved. 
Throughout this work, like atoms are taken to be non-interacting,
implying $a_{\alpha \alpha}=0$ for $\alpha= 1, \cdots, \chi$.

In the most general case,
the Hamiltonian given in Eq.~(\ref{eq_ham}) 
has $\chi$ different 
$m_{\alpha}$, $\omega_{\alpha}$ and $N_{\alpha}$ 
($\alpha=1,\cdots,\chi$),
and $\chi(\chi-1)/2$
different $V_{\alpha \beta}$ ($\alpha,\beta=1,\cdots,\chi$
and $\alpha \ne \beta$),
resulting in a
tremendously large parameter space.
To 
reduce the parameter space, we
first consider the case where
all $m_{\alpha}$,
all $\omega_{\alpha}$,
all $N_{\alpha}$, and
all $a_{\alpha \beta}$ ($\alpha \ne \beta$) are equal.
A four-component gas of this type could, e.g., be realized 
by equally populating and trapping the four spin states
of a fermionic atom whose ground state has vanishing 
total
electronic angular momentum 
$J$ but a non-vanishing
nuclear spin $I$ of $3/2$.
In this case, 
the scattering lengths between the 
different spin substates are equal and $s$-wave scattering between
two atoms in the same spin substate are forbidden by
symmetry. 

Before solving the Schr\"odinger equation for the
Hamiltonian given in Eq.~(\ref{eq_ham}),
we present a simple counting analysis.
The number $N_{att}$ of attractive pair interactions $V_{\alpha \beta}$,
where again $\alpha \ne \beta$,
is given by
\begin{eqnarray}
N_{att}=\frac{N^2}{2} \frac{\chi - 1}{\chi},
\end{eqnarray}
and the number $N_{rep}$ of
effectively repulsive interactions, i.e., the number of like fermion pairs, 
by
\begin{eqnarray}
N_{rep}=\frac{N}{2} \frac{N-\chi}{\chi}.
\end{eqnarray}
Table~\ref{tab1} summarizes the values of $N_{att}$,
\begin{table}
\caption{\label{tab1} Number $N_{att}$ of attractive interactions,
number $N_{rep}$ of effectively repulsive interactions and ratio 
$N_{rep}/N_{att}$ for finite and infinite $N$
for a $\chi$-component Fermi gas 
($\chi=2$ through $4$) in which all interspecies 
interactions are equal (or resonant).}
\begin{ruledtabular}
\begin{tabular}{l|ccc}
  & $\chi=2$ & $\chi=3$ & $\chi=4$  \\ \hline
 $N_{att}$ & $\frac{1}{4}N^2$ & $\frac{1}{3}N^2$ & $\frac{3}{8}N^2$ \\
 $N_{rep}$ & $\frac{N}{2}\left( \frac{N}{2}-1 \right)$ & $\frac{N}{2}\left( \frac{N}{3}-1 \right)$ & $\frac{N}{2}\left( \frac{N}{4}-1 \right)$ \\
$N_{rep}/N_{att}$ ($N$ finite) & $\frac{N-2}{N}$ &$\frac{N-3}{2N}$ &$\frac{N-4}{3N}$ \\
$N_{rep}/N_{att}$ ($N \rightarrow \infty$) & $1$ &$\frac{1}{2}$ &$\frac{1}{3}$ \\
\end{tabular}
\end{ruledtabular}
\end{table}
$N_{rep}$ and $N_{rep}/N_{att}$ for $\chi=2$, $3$ and $4$.
The ratio $N_{rep}/N_{att}$ (reported
in the fourth and fifth row of Table~\ref{tab1} for finite
and infinite $N$, respectively)
decreases with increasing $\chi$ and approaches
$1/(\chi-1)$ in the large $N$ limit,
indicating that the Fermi pressure becomes less
important compared to the interspecies interactions as $\chi$
increases.
Another interesting scenario arises
when each component is occupied 
by exactly one fermion, i.e., when $\chi=N$. In this case,
no effectively repulsive interactions exist,
i.e., $N_{rep}/N_{att}=0$, and the system's ground state is the same as that of
the corresponding $N$-boson system. As 
pointed out already in the introduction,
Bose gases 
with negative scattering length
become
unstable in the 
limit that the absolute value of the scattering length becomes large.
This, together with the fact that two-component Fermi gases are 
stable for all scattering lengths, suggests that
there exists a critical $\chi$-value beyond which
multi-component Fermi gases with large negative interspecies
scattering length
are unstable. Sections~\ref{sec_hyper} and \ref{sec_mc} show
that $\chi_{cr}=3$.

In addition to Fermi gases in which all interspecies interactions 
$V_{\alpha \beta}$ ($\alpha,\beta=1,\cdots,\chi$ and $\alpha \ne \beta$) are
equal,
we consider the scenario in which only a subset of interspecies interactions
are ``turned on''. 
In particular, we consider $\chi$-component Fermi gases with
$\chi-1$ equal and non-zero (or resonant) $a_{\alpha \beta}$.
For the three-component system, we take
$a_{12}=a_{23}$ and $a_{31}=0$.
The number of attractive
interactions is in this case by a factor of 2/3 smaller than in the case
with three
resonant interactions, thus increasing the ratio of $N_{rep}/N_{att}$
from 1/2 to 3/4 in the large $N$ limit. 
For the four-component system, 
two
different 
``non-trivial'' 
possibilities for turning on only
$\chi-1$ 
interactions 
exist:
(i) $a_{12}=a_{13}=a_{14}$ and $a_{23}=a_{34}=a_{24}=0$, and
(ii) $a_{12}=a_{23}=a_{34}$ and $a_{13}=a_{14}=a_{24}=0$
[the configuration
$a_{12}=a_{23}=a_{34}$ and $a_{13}=a_{14}=a_{24}=0$, e.g.,
is equivalent to (i);
the configuration 
$a_{12}=a_{23}=a_{31}$ and $a_{14}=a_{24}=a_{34}=0$, e.g.,
is trivial in the sense that it can be broken up into a three-component system
with all resonant interactions and a single non-interacting 
component].
For the non-trivial configurations,
the number of attractive
interactions is 
half as large for
the case of three resonant interactions
as 
in the case
of six
resonant interactions, thus increasing the ratio of $N_{rep}/N_{att}$
from 1/3 to 2/3 in the large $N$ limit. 
This counting analysis indicates that the values of the ratio
$N_{rep}/N_{att}$ for large $N$
for three- and four-component gases with $\chi-1$ 
resonant $a_{\alpha \beta}$ are between those for 
two- and three-component Fermi gases
with $\chi(\chi-1)/2$ resonant $a_{\alpha \beta}$.
Thus, the question arises whether
multi-component Fermi gases
with $\chi-1$ resonant
interactions are stable for all scattering lengths
and $N$ values, or whether they become unstable for a certain critical 
parameter combination.

To analyze the stability of Fermi gases quantitatively,
one may attempt to 
describe the interspecies atom-atom interactions by a zero-range Fermi
pseudopotential 
$V_{\delta}(\vec{r})$~\cite{ferm34},
\begin{eqnarray}
\label{eq_pp}
V_{\delta}(\vec{r}) =
\frac{2 \pi \hbar^2 a_s}{\mu}  \delta(\vec{r}),
\end{eqnarray}
which is directly proportional to the $s$-wave scattering length $a_s$.
Here, $\mu$ denotes the reduced mass of the two interacting atoms.
This pseudopotential has been employed
successfully to
predict many properties of 
dilute Bose gases.
For example, the interaction 
potential
given in Eq.~(\ref{eq_pp})  
together with a Hartree wave function correctly predicts
that trapped Bose gases
with negative $a_s$
become
unstable if~\cite{dodd96,esry97} 
\begin{eqnarray}
\label{eq_critbose}
(N-1) \frac{a_s}{a_{ho}} \lesssim -0.575,
\end{eqnarray}
where
$a_{ho}$ denotes the oscillator length, 
$a_{ho} = \sqrt{\hbar/(2 \mu \omega)}$.
In general, the true atom-atom 
interaction can be replaced in the long-wavelength limit 
by the pseudopotential given in Eq.~(\ref{eq_pp})
provided the system is dilute, i.e., if $n(0)|a_s|^3 \ll 1$,
where $n(0)$ denotes the peak density. 
Assuming $N$ is not too small, one finds that
$n(0)|a_s|^3$ is much smaller than one when
$(N-1)a_s$ equals $-0.575a_{ho}$;
consequently,
the pseudopotential predicts the instability of dilute
Bose gases 
correctly.

Applied to two-component 
Fermi gases, the bare pseudopotential employed within a hyperspherical
framework predicts that the
system becomes unstable if~\cite{ritt06}
\begin{eqnarray}
\label{eq_critfermi}
k_F(0) a_s \lesssim -1.22,
\end{eqnarray}
where $k_F(0)$ denotes the noninteracting Fermi wave
vector at the trap center. 
(A mean-field analysis predicts a slightly more negative critical
value 
of $k_F(0) a_s \lesssim -\pi/2$~\cite{gehm03,stoo96,houb97}.) 
However,
at the critical parameter combination
$k_F(0) a_s =-1.22$,
the diluteness criterium,
which can be written 
as $(k_F(0) |a_s|)^3 \ll 1$ for fermions,
is violated and the 
instability prediction, Eq.~(\ref{eq_critfermi}),
cannot be trusted.
Indeed, Eq.~(\ref{eq_critfermi})
disagrees with the experimental finding that two-component
Fermi gases are stable~\cite{zwie03,grei03,stre03,bour04,kina04,bart04}.
To resolve this disagreement,
two independent studies
recently introduced density-dependent 
``renormalization schemes'' of the scattering length $a_s$ entering into 
Eq.~(\ref{eq_pp})~\cite{freg06,stec07}.
Using these 
``renormalization schemes'',
the (modified)
density-dependent 
pseudopotential
can be applied to describe strongly-interacting Fermi gases
and predicts,
in agreement with 
experimental~\cite{zwie03,grei03,stre03,bour04,kina04,bart04}
and other theoretical~\cite{bake99,heis01,carl03,astr04c,chan04} works,
that two-component Fermi gases are stable even in the 
strongly-interacting unitary regime.
The fact that the Fermi pseudopotential has to be modified
when applied to 
fermions was already suggested earlier (see, e.g.,
Refs.~\cite{fedo01,gehm03}),
and is well known in the nuclear physics community
(see, e.g., Refs.~\cite{chab98,faya00} and references therein).

The instability prediction given in Eq.~(\ref{eq_critfermi})
for two-component Fermi gases can be readily generalized
to multi-component Fermi gases with $\chi(\chi-1)/2$ and with $\chi-1$
resonant interactions (in the case of $\chi-1$ resonant interactions, we
consider the non-trivial scenario introduced above, in which
none of the components can be separated off). 
The right hand side of Eq.~(\ref{eq_critfermi})
has to be multiplied by 
$1/(\chi - 1)$ in the former case, and by
$\chi/[2(\chi-1)]$ in the latter case.
Since the diluteness criterium is fullfilled approximately
in the all resonant case at the predicted collapse point 
[e.g., $(k_F(0) a_s)^3 \approx -0.23$ and $-0.067$ for $\chi=3$ and 4,
respectively],
the prediction that multi-component Fermi gases with all
resonant interactions become unstable for a finite $|a_s|$,
derived using the 
``bare'' interaction given in Eq.~(\ref{eq_pp}),
may in fact be correct. Sections~\ref{sec_hyper} and \ref{sec_mc}
confirm this.
In the case with $\chi-1$ resonant interactions,
the bare Fermi pseudopotential, Eq.~(\ref{eq_pp}), predicts that
the collapse occurs at
$(k_F(0) a_s)^3 \approx -0.76$ and $-0.54$ for $\chi=3$ and 4, 
respectively. This suggests that the bare 
pseudopotential cannot be used and that the instability prediction
derived using it may not be correct
(see Sec.~\ref{sec_mc} for a many-body analysis).

\section{Hyperspherical framework}
\label{sec_hyper}

\subsection{$N$-particle system}
\label{sec_hyper_n}
This section investigates the 
stability of three- and four-component Fermi gases
within a hyperspherical framework~\cite{bohn98,sore01,ritt06,ritt07}.
Throughout this section, we assume that all angular trapping frequencies
are equal, i.e., 
$\omega_{\alpha}=\omega$ for $\alpha=1,\cdots,\chi$.
To gain insight into the system's behavior,
we employ hyperspherical coordinates:
the 
$3N$ coordinates are divided into a hyperradius $R$ and $3N-1$ hyperangles,
collectively denoted by $\Omega$~\cite{aver89}. 
In the following, only the definition
of the hyperradius $R$, which can be thought of as a measure of the size
of the system, is needed,
\begin{eqnarray}
\label{eq_hyper} 
R= 
\left(
\frac{1}{M}\sum_{\alpha=1}^{\chi}\sum_{i=1}^{N_{\alpha}} 
m_{\alpha} \vec{r}_{\alpha i}^2
\right)^{\frac{1}{2}}.
\end{eqnarray}
As before, the position vectors
$\vec{r}_{\alpha i}$ are measured with respect
to the center of the trap.
In Eq.~(\ref{eq_hyper}), 
$M$ denotes the total mass of the system,
i.e., $M = \sum_{\alpha=1}^{\chi}\sum_{i=1}^{N_{\alpha}} 
m_{\alpha}$.
Using these coordinates, the many-body
wave function $\Psi(R,\Omega)$  can be expanded 
in terms of a set of $\Omega$-dependent channel functions 
$\Phi_{\nu}(\Omega;R)$, which
depend
parametrically on $R$,
and $R$-dependent weight functions $F_{\nu n}(R)$,
\begin{eqnarray}
\label{eq_wavefct}
\Psi(R,\Omega) = \sum_{\nu n} R^{(1-3N)/2} F_{\nu n}(R) \Phi_{\nu}(\Omega;R).
\end{eqnarray}

In the adiabatic hyperspherical approximation~\cite{mace68,klar78,star79,lin95,bohn98}, 
which neglects
off-diagonal coupling elements and additionally restricts the sum
in Eq.~(\ref{eq_wavefct}) to a single term,
the solution of the
many-body Schr\"odinger equation reduces to determining
an $R$-dependent 
effective potential $V_{\nu}(R)$, 
which includes part of the kinetic energy as well
as the effects of the two-body interactions,
and then solving an effective 
one-dimensional radial Schr\"odinger equation in the hyperradius $R$,
\begin{eqnarray}
\label{eq_hyper1}
\left[ -\frac{\hbar^2}{2M} \frac{\partial^2}{\partial R^2}
+ V_{\nu}(R) +V_{trap}(R) 
\right]
F_{\nu n}(R) = \nonumber \\
E_{\nu n} F_{\nu n}(R),
\end{eqnarray}
where
\begin{eqnarray}
V_{trap}(R)=\frac{1}{2} M \omega^2 R^2.
\end{eqnarray} 
In general, the calculation of the effective potentials $V_{\nu}(R)$
is, at least for many-body systems, just as hard as solving the many-body
Schr\"odinger equation itself.
In certain circumstances, however, the effective potentials
or their functional form can be obtained more easily.

For weakly-interacting dilute equal-mass two-component Fermi gases, e.g.,
the effective hyperradial potential $V_0(R)$ consists of two 
parts~\cite{ritt06,ritt07}:
The first part, which is proportional to $c_{kin}/R^2$,  
arises from the kinetic energy operator, and the second part,
which is proportional to $c_{int}/R^3$, accounts for the
particle-particle interactions. 
$c_{kin}$ is positive, and $c_{int}$ is 
directly proportional to the $s$-wave scattering length 
$a_s$~\cite{ritt06}.
If $|a_s|$ is sufficiently small ($a_s<0$),
the small $R$ region, where the attractive $c_{int}/R^3$ term dominates
over the repulsive $c_{kin}/R^2$ term, is separated by a ``barrier'' from the 
$R$ region where the repulsive $c_{kin}/R^2$ term dominates
over the attractive $c_{int}/R^3$ term. This barrier prevents the
Fermi gas from collapsing to a cluster-like many-body bound state,
thus explaining the stability of weakly-interacting dilute equal-mass
two-component Fermi gases (see also Ref.~\cite{freg06}).

Analytical expressions
for $V_{\nu}(R)$ for equal-mass two-component 
Fermi gases have also been obtained in the strongly-interacting
limit~\cite{wern06,ritt07}.
For infinitely large interspecies
scattering lengths $a_{\alpha \beta}$ ($\alpha \ne \beta$),
the adiabatic approximation
introduced above is--- for a class
of states that arise assuming 
boundary conditions consistent with contact interactions
when the distance between each pair of particles goes to zero~\cite{wern06}, 
referred to as universal states
in the following--- exact and
the universal effective potentials $V_{\nu}(R)$
are given by
\begin{eqnarray}
\label{eq_hyper2}
V_{\nu}(R)= \frac{\hbar^2 (1+C_{\nu})}{2 M R^2}.
\end{eqnarray}
In Eq.~(\ref{eq_hyper2}), the $C_{\nu}$ denote 
$R$-independent constants that
arise
from the integration  
over
the $3N-1$ hyperangles. 
The functional form given in Eq.~(\ref{eq_hyper2}) has also been derived
by 
explicitly solving the many-body 
Schr\"odinger equation
for a class of
variational wave functions
using hyperspherical
coordinates~\cite{ritt07}. 
Using dimensional arguments,
Eq.~(\ref{eq_hyper2}) can be understood as follows:
The term of $V_{\nu}(R)$
that accounts for the particle-particle interactions has---
because of the absence of any other length scale in the problem--- 
to scale in the
same way with $R$ as the $1/R^2$ 
term that accounts for the kinetic energy contribution.
The total universal effective hyperradial
potential curves at unitarity thus have a simple functional form:
$V_{\nu}(R)$ dominates at small $R$ and approaches plus or minus infinity
in the $R \rightarrow 0$ limit,
depending on whether the quantity $1+C_{\nu}$ is positive or negative,
and $V_{trap}(R)$ dominates at large $R$.

For a two-component unitary Fermi gas, a number of $C_{\nu}$
values are known. For $N \le 6$, selected $C_{\nu}$ 
with $\nu \le 2$ have been 
obtained using a correlated Gaussian (CG) basis set expansion-type
approach~\cite{blum07}.
For $N \le 30$, 
upper bounds for the $C_0$ have been obtained by the FN-DMC 
method~\cite{blum07}.

\begin{figure}
\vspace*{.075in} 
\includegraphics[angle=270,width=80mm]{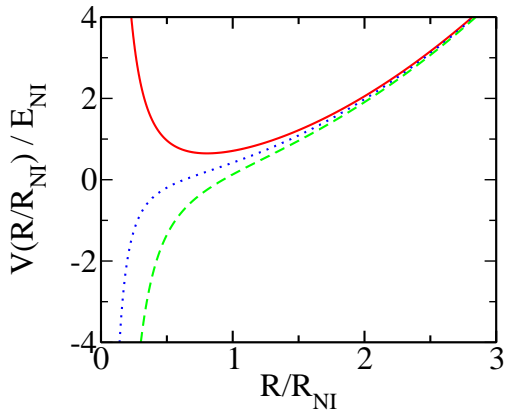}
\caption{
(Color online)
Hyperradial potential curve $V_0(R) + V_{trap}(R)$
as a function of the hyperradius $R$ for $\chi=2$ 
(solid line), $\chi=3$ (dotted line) and $\chi=4$
(dashed line)
in the large $N$ limit. All interactions between unlike atoms 
are characterized by an infinite
scattering length, and the coefficient $C_0$ is taken to be 
$(\chi-1)\beta$
with $\beta=-0.58$.
Both length and energy are scaled by the corresponding values of
the non-interacting system 
(see text).
}
\label{fig1}
\end{figure}

In the large $N$ limit,
the value of $C_0$ for equal-mass two-component 
Fermi gases at unitarity
can be related to the universal parameter $\beta$ of the homogeneous system
using the hyperspherical framework of Ref.~\cite{ritt07}
that employs a 
density-dependent scattering
length~\cite{stec07}, leading to
$C_0=\beta$~\cite{ritt07}.
Alternatively, this relationship can be derived
by applying the local 
density approximation to the trapped system 
(see, e.g., Ref.~\cite{blum07}).
It
is generally believed that the most accurate value for $\beta$ has been
obtained by the FN-DMC method~\cite{astr04c,carl05}, $\beta = -0.58$.
Since $1+\beta>0$,
the universal hyperradial potential
curve $V_0(R)+V_{trap}(R)$ 
for large $N$ (shown by a solid line in Fig.~\ref{fig1} using $\beta=-0.58$)
is repulsive 
at small $R$, 
preventing the collapse of the two-component 
unitary Fermi gas towards cluster-like bound states. 
The hyperspherical potential curve picture reveals an intuitive way
of understanding the stability of the energetically lowest-lying
gas-like state of two-component unitary Fermi gases.

In Fig.~\ref{fig1}, the 
hyperradial potential $V(R)$ 
and the
hyperradius $R$ 
are scaled
by $E_{NI}$
and  $R_{NI}$, 
respectively, where
$E_{NI}$ denotes the energy of the non-interacting $\chi$-component Fermi
system and
$R_{NI}$ the square root of
the expectation value of $R^2$ calculated for the non-interacting 
$\chi$-component Fermi system,
i.e., $R_{NI}= \sqrt{\langle R^2 \rangle_{NI}}$~\cite{ritt06,ritt07}.
Using the virial theorem~\cite{thom05,wern06}, $R_{NI}$ 
can be related to $E_{NI}$,
\begin{eqnarray}
R_{NI}= \sqrt{\frac{\hbar}{M \omega}} \sqrt{\frac{E_{NI}}{\hbar \omega}}.
\end{eqnarray}
In Fig.~\ref{fig1}, $E_{NI}$ and $R_{NI}$ have been evaluated by
writing $E_{NI}$ as $(\lambda + 3N/2) \hbar \omega$ and 
using the leading order of $\lambda$ for large closed-shell systems,
$\lambda = (3N)^{4/3}/4$.
We note that
the scaled quantity $R/R_{NI}$ remains finite
in the limit that $N$ goes to infinity, allowing the
thermodynamic limit to be taken.

For all-resonant
three- and four-component gases
at unitarity, the functional form given in Eq.~(\ref{eq_hyper2})
remains valid for universal states~\cite{wern06,ritt07}.
For these systems, the values of $C_{\nu}$
are not known
(the only exception being the $N=3$ system~\cite{dinc05,wern06b}, 
see below).
It has been argued~\cite{ritt07}, parametrizing $V_{\alpha \beta}$ 
by a 
density-dependent 
zero-range two-body potential~\cite{stec07} and neglecting the existence 
of weakly-bound trimer or tetramer states, that the coefficient 
$C_0$ is in the large $N$ limit determined by the
parameter $\beta$ of the unitary two-component Fermi gas, i.e., 
$C_0=(\chi-1) \beta$.
Due to the lack of benchmark results for 
all-resonant three- and four-component
Fermi gases, there is some ambiguity in how
the simplest adaptation of the renormalization scheme, originally designed to
describe the physics of two-component Fermi gases~\cite{stec07},
is applied to three- and four-component Fermi gases,
thus introducing some uncertainty about the exact values of $C_{0}$
that describe unitary three- and four-component Fermi gases.
Using 
$C_0=(\chi-1) \beta$ and $\beta=-0.58$~\cite{astr04c,carl05},
$(1+C_0)$ is negative for three- and four-component gases, 
$(1+C_0)=-0.16$ and $(1+C_0)=-0.74$
for $\chi=3$ and 4, respectively~\cite{footnoterenorm}.
The resulting hyperradial potential curves $V_{0}(R)+V_{trap}(R)$
are shown in Fig.~\ref{fig1} in the large $N$ limit by
dotted and dashed lines for $\chi=3$ and 4, respectively.
The attractive small-$R$ behavior is due to the
negative values of $(1+C_0)$, and suggests that unitary three- and
four-component Fermi gases with all resonant interations are unstable.
Note that the coefficient $C_0$ 
and $N_{att}/N_{rep}$ both scale as $1-\chi$ with
$\chi$, reflecting the fact that $V_{0}(R)$ is in part determined
by the pair interactions.

Interestingly, both the bare pseudo-potential (see the discussion
towards the end of Sec.~\ref{sec_system}) and the 
density-dependent
pseudo-potential (used in this section within the 
hyperspherical framework) predict that $\chi$-component
Fermi gases, $\chi>2$, with 
all resonant interactions become
unstable for a finite value of the interspecies scattering length.
The critical value predicted by the bare interaction is
presumably not negative enough while that for the 
density-dependent
interaction
would presumably be too negative (we expect
that the renormalization
scheme originally developed for the two-component Fermi gas
``over-renormalizes'' the scattering length).

\begin{figure}
\vspace*{.075in} 
\includegraphics[angle=270,width=80mm]{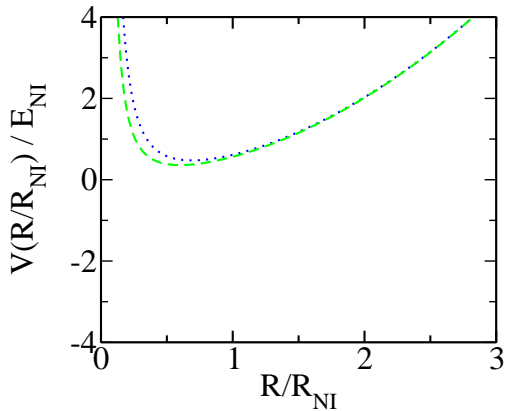}
\caption{
(Color online)
Hyperradial potential curve $V_0(R) + V_{trap}(R)$
as a function of the hyperradius $R$ for $\chi=3$ (dotted line) and $\chi=4$
(dashed line) in the large $N$ limit for $\chi-1$ resonant interactions 
(the scattering lengths $a_{1\beta}$, $\beta=2,\cdots,\chi$, are infinite 
and all other scattering lengths are zero).
The coefficient $C_0$ is taken to be $\frac{2}{\chi}(\chi-1)\beta$
with $\beta=-0.58$.
Both length and energy are scaled by the corresponding values of
the non-interacting system 
(see text).
The plotting range of the axis is the same as in 
Fig.~\protect\ref{fig1} for ease
of comparison.
}
\label{fig2}
\end{figure}

The analysis above can be extended to the case where only 
$\chi-1$ interspecies scattering lengths 
are infinite; 
as before, we focuss on what we defined in Sec.~\ref{sec_system} as 
``non-trivial'' scenarios.
In the case of $\chi-1$ resonant interactions,
the bare pseudopotential is expected to fail (see
Sec.~\ref{sec_system}).
Using, just as above, density-dependent
zero-range two-body interactions to describe the $V_{\alpha \beta}$
with non-zero $a_{\alpha \beta}$,
the coefficient
$C_0$ becomes in the large $N$ limit $\frac{2}{\chi}(\chi-1) \beta$;
just as in the all-resonant case, 
$C_0$ scales with $\chi$ in the same way
as $N_{att}/N_{rep}$. 
Using $\beta=-0.58$, 
$(1+C_0)$ is positive
for $\chi=3$ and $4$ (the corresponding potential curves are shown
in Fig.~\ref{fig2});
the repulsive small-$R$ behavior suggests that
three- and four-component
unitary Fermi gases with $\chi-1$ resonant interactions
are stable. 
We note that the analysis in hyperspherical coordinates,
which employs
density-dependent
interactions, has two short-comings:
i)
The derivation of the effective potential curves assumes
the existence of a single length scale, the hyperradius $R$;
since
some of the interspecies scattering length are zero while 
others are resonant, a
more accurate description would presumably introduce an 
additional length
scale and be based on a ``hyperradial vector'' as opposed to
a single hyperradius.
ii) The derivation
neglects the 
possible existence of weakly-bound trimer (see the next section)
or tetramer states,
or said differently, the existence of non-universal states.
Section~\ref{sec_mcresults} 
shows that the behaviors of multi-component Fermi gases with
$\chi>2$ may differ depending on whether or not such states exist.

\subsection{Three-particle system}
\label{sec_hyper_3}
This section discusses the behaviors of
three distinguishable fermions interacting through
zero-range potentials with equal masses and equal trapping frequencies.
For these systems, 
the effective 
hyperradial potential curves $V_{\nu}(R)$ can be obtained analytically
for any combination of scattering lengths $a_{12}$, $a_{23}$
and $a_{31}$~\cite{niel99,niel01}. 

\begin{figure}
\vspace*{.075in} 
\includegraphics[angle=270,width=80mm]{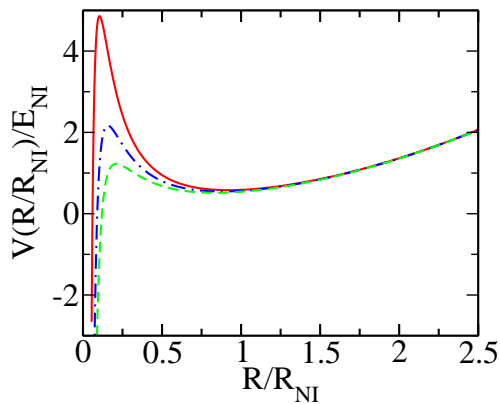}
\caption{
(Color online)
Solid, dash-dotted and dashed lines show the
hyperradial potential curve $V_0(R)+V_{trap}(R)$
as a function of the hyperradius $R$, scaled by $R_{NI}$,
for three distinguishable fermions interacting through
zero-range interactions with $a_s=-0.1a_{ho}$,
$-0.15a_{ho}$ and $-0.2 a_{ho}$, respectively.
Note that $R$ is defined without the center-of-mass motion,
implying $E_{NI}=3 \hbar \omega$.
}
\label{hyper3}
\end{figure}

Figure~\ref{hyper3} shows $V_0(R)+ V_{trap}(R)$
for the three-particle system 
as a function of $R$ for three different 
negative scattering lengths $a_s$, i.e., $a_s=-0.1a_{ho}$, $-0.15a_{ho}$
and $-0.2 a_{ho}$ 
(here, $a_s=a_{\alpha \beta}$ with $\alpha \ne \beta$)
as a function of the hyperradius $R$.
Throughout this section, 
the hyperradius $R$ is defined without the center-of-mass motion
and scaled by the hyperradius $R_{NI}$
of the non-interacting system without
the center-of-mass motion.
For the discussion that follows, the difference between the hyperradius
defined without the center-of-mass 
and that defined in Eq.~(\ref{eq_hyper}) is irrelevant.
For the scattering lengths shown, 
the hyperradial potentials show a barrier around
$R\approx 0.2R_{NI}$ (i.e., $R \approx 0.3 a_{ho}$), which separates the
large $R$ region where gas-like states live from
the small $R$ region. As $a_s$ becomes more negative, the barrier
height decreases and moves to slightly larger $R$
values. The critical scattering length
$a_c$ at which the barrier disappears ($a_c \approx -0.39 a_{ho}$)
provides a first estimate for when the three-body
system becomes ``unstable'' against the formation
of tightly-bound trimer states with negative energy,
assuming such states exist;
the conditions for the existence of negative energy states are
discussed below.
A more accurate estimate would account for the zero-point energy of the
quantum system, resulting in a somewhat less negative
critical scattering length that is in fair agreement
with the GP prediction given in Eq.~(\ref{eq_critbose})
(see Ref.~\cite{bohn98} for a discussion of the $N$-boson system). 
The lifetime
of the gas-like three-body state can be estimated by
calculating the tunneling rate through the potential
barrier as a function of $a_s$~\cite{bohn98}. 
When the barrier height is large compared to the energy
of the non-interacting system (small $|a_s|$), the lifetime of
the gas-like trimer state is
much larger than $1/\omega$;
in this case, the system can be considered stable
(it is, in fact, metastable). 
The tunneling rate is appreciable only when the barrier height
becomes comparable to the energy of the non-interacting system,
i.e., when $a_s$ approaches $a_c$.

To obtain the energy spectrum in the adiabatic
approximation, we solve the 
one-dimensional radial Schr\"odinger equation
[Eq.~(\ref{eq_hyper1})]. To this end,
we 
add a term $V_{SR}(R)$
that is repulsive for $R \lesssim R_c$ 
and negligibly small for $R \gtrsim R_c $ to $V_{\nu}(R)$,
$V_{SR}(R)= 2(1-\sqrt{3}{\mbox{tanh}}^2(R/R_c))/(2 m R^2)$.
$V_{SR}(R)$ 
cures the divergence of the $V_{\nu}(R)$
at small $R$, which is related
to the Thomas collapse~\cite{thom35}.
When the barrier height is large
compared to the energy of the non-interacting system
and 
the short-range length scale
$R_c$ is chosen sufficiently small,
this ``ad-hoc'' regularization should affect the eigenenergies of the
gas-like states at most weakly. The eigenenergies of molecular-like states
(states living in the small $R$ region), however, 
should depend comparatively
strongly on the particular value of $R_c$.
Thus, some properties of the system are expected to be non-universal.
Indeed, our calculations for different $R_c$ 
show that the first state with negative energy appears 
when the scattering length $a_s$ is approximately equal to 
$-8.4 R_c$.
For $R_c=0.001a_{ho}$, e.g., the first negative energy
state appears 
for 
$a_s \approx -0.0084 a_{ho}$,
a value much
less negative than $a_c$. At this scattering length, the barrier height
is large compared to the energy of the non-interacting system and the
gas-like system is--- because of the existence of a negative trimer
state--- metastable.
If $R_c$ is chosen so that the absolute value of
the scattering length at which the first bound
state appears is comparable to or larger than $|a_c|$, then the 
gas-like state would be the true ground state of the system
for $|a_s| < |a_c|$. 
For realistic dilute alkali gases with all-resonant interactions, 
we expect that three-body
bound states appear for scattering lengths less negative than $a_c$;
consequently, trimer as well as larger three-component systems 
with $a_s \gtrsim a_c$ are expected to be metastable.
The picture developed here within the adiabatic approximation
remains qualitatively valid if the coupling between channels is included.

In addition to the all-resonant three-particle system, we consider the 
trapped three-particle system with two resonant interactions
($a_s=a_{12}=a_{23}$ and $a_{31}=0$).
In this case, the hyperradial
potential barrier disappears at a 
more negative scattering length than in the all-resonant case, i.e.,
at $a_c \approx -1.00 a_{ho}$.
An analysis of the tunneling
probability suggests that the lifetime of the 
two-resonant system is, considering the same $a_s$, enhanced
by a factor of about ten compared to the all-resonant
system. A lifetime enhancement is
expected since the ratio $N_{att}/N_{rep}$ 
(see Sec.~\ref{sec_system}) is smaller for
the three-component system with two resonant interactions than
for that with three resonant interactions.

Calculating the trimer energies in the adiabatic approximation, we
find that the first three-body state with negative energy
appears at $a_s \approx -2500 R_c $.
For example, a negative energy state appears for
$a_s \approx -2.5 a_{ho}$ if $R_c=0.001a_{ho}$ and for
$a_s \approx -0.25 a_{ho}$ if $R_c=0.0001a_{ho}$. 
In the latter case, the disappearing of the hyperradial potential
barrier is accompanied by 
a ``collapse'' of the metastable gas-like state to
a cluster-like state. In the former case, in contrast, the disappearing of the
hyperradial potential barrier is
accompanied by the lowest-lying gas-like
state evolving smoothly to a cluster-like state with negative energy.
Although the 
short-range
parameter $R_c$ cannot be straightforwardly related to
the parameters of typical atom-atom potentials, it seems plausible that 
realistic three-component Fermi systems with two resonant interactions
could fall into either of the 
two categories
discussed here.
Section~\ref{sec_mc} investigates 
the behaviors of three-component many-body systems and 
attempts to determine how the system's behaviors depend on the 
range of the underlying finite-range two-body potential
(this range can, roughly speaking,
be connected with the 
short-range parameter 
$R_c$).

We note that the negative scattering length $a_s$ at which the first 
weakly-bound trimer state appears
(assuming no deeply-bound states exist), can be estimated using the 
following ``rule of thumb''~\cite{efim70,efim71,efim73}:
$|a_s| = R_c \exp(\pi / s_0)$, 
where $s_0$ is the coefficient that determines
the energy spectrum of the lowest hyperradial potential curve at unitarity
($s_0=1.006$ for $a_s=a_{12}=a_{23}=a_{31}$,
and $s_0=0.413$ for $a_s=a_{12}=a_{23}$ and $a_{31}=0$~\cite{dinc05}).
This rule of thumb predicts a negative scattering length
at which the first weakly-bound trimer state appears 
of 
$-23 R_c$ and $-2000 R_c$ 
for three and two resonant interactions, respectively.
These values are 
to be compared with 
$-8.4 R_c$ and $-2500R_c$ 
found numerically (see above).

\section{Monte Carlo treatment}
\label{sec_mc}
This section discusses the solutions of the many-body 
Schr\"odinger equation for the Hamiltonian given in 
Eq.~(\ref{eq_ham}) obtained by Monte Carlo techniques
as a function of the scattering length.
Section~\ref{sec_mcdetail} introduces the VMC and FN-DMC methods,
and Sec.~\ref{sec_mcresults} presents our results.

\subsection{Variational and fixed-node diffusion Monte Carlo method}
\label{sec_mcdetail}
Using the VMC and FN-DMC methods~\cite{hamm94},
we determine solutions of the many-body Schr\"odinger equation
for two different 
purely
attractive short-range model potentials $V_{\alpha \beta}$:
a square well interaction
potential $V_{\alpha \beta}^{SW}$,
\begin{eqnarray}
V_{\alpha \beta}^{SW}(r) = 
\left\{ \begin{array}{ll} -V_{\alpha \beta,0} & \mbox{for } r < R_{\alpha \beta, 0} \\ 
0 & \mbox{for } r > R_{\alpha \beta, 0}, 
\end{array} \right.
\end{eqnarray}
and a Gaussian interaction
potential $V_{\alpha \beta}^G$,
\begin{eqnarray}
V_{\alpha \beta}^{G}(r)=-V_{\alpha \beta,0}
\exp \left(-\frac{r^2}{2 R_{\alpha \beta, 0}^2} \right).
\end{eqnarray} 
Both potentials depend on 
a depth $V_{\alpha \beta, 0}$, $V_{\alpha \beta,0} \ge 0$,
and a range $R_{\alpha \beta,0}$. 
In our calculations, we set all $R_{\alpha \beta,0}$
to the same value, $R_{\alpha \beta,0} \ll a_{ho}$,
and adjust the depths $V_{\alpha \beta,0}$
until the $s$-wave scattering lengths $a_{\alpha \beta}$
assume the desired values. 
In Sec.~\ref{sec_mcresults},
$V_{\alpha \beta,0}$ and $R_{\alpha \beta,0}$ are chosen so
that the potential $V_{\alpha \beta}$ supports no 
two-body $s$-wave bound state and so that
$a_{\alpha \beta} \le 0$.
The use of two different functional forms for
the interaction potentials $V_{\alpha \beta}$
allows for exploring the dependence of the results
on the details of the short-range
physics.

We now discuss how the solutions of the
many-body Schr\"odinger equation 
are obtained by the
variational Monte Carlo method.
The functional form of the variational 
wave function $\psi_T$ is chosen on physical
grounds (see below), and $\psi_T$ is parametrized in terms of
a set of parameters $\vec{p}$, which are optimized so as to
minimize the energy expectation value
$E(\vec{p})$~\cite{hamm94},
\begin{eqnarray}
\label{eq_vmc}
E(\vec{p}) = \frac{\langle \psi_T(\vec{p}) | H | \psi_T(\vec{p}) \rangle}
{\langle \psi_T(\vec{p}) | \psi_T(\vec{p}) \rangle}.
\end{eqnarray}
Since $\psi_T$ depends in general on the $3N$ coordinates of the system,
the integration on the right hand side of Eq.~(\ref{eq_vmc}) is high-dimensional;
this implies that standard techniques such as the Simpson rule~\cite{pres92}, 
which scale,
roughly speaking, exponentially with the number of degrees of freedom
cannot be used for its evaluation.
Instead, we evaluate the high-dimensional integral by Monte Carlo techniques
using standard
Metropolis sampling~\cite{metr53}. 
The stochastic evaluation of the integral introduces a statistical
uncertainty,
which can be reduced
by increasing the number of samples $K$ included in the evaluation
of the expectation value (the errorbar decreases
as $1/\sqrt{K}$ with increasing $K$).
We denote the energy expectation value obtained by the VMC method by 
$E_{VMC}$.

The variational wave function $\psi_T$ for multi-component Fermi gases
is written as (see 
Refs.~\cite{carl03,astr04c,chan04,jaur07,chan07,stec07b,blum07} for
Monte Carlo studies of two-component systems and 
Ref.~\cite{chan06} for a Monte Carlo study of three-component systems)
\begin{eqnarray}
\label{eq_trial}
\psi_T =
\prod_{\alpha < \beta}^{\chi}
\prod_{i=1}^{N_{\alpha}} \prod_{j=1}^{N_{\beta}} 
f_{\alpha \beta} (| \vec{r}_{\alpha i} - \vec{r}_{\beta j}| ) 
\times \prod_{\alpha=1}^{\chi} \prod_{i = 1}^{N_{\alpha}} 
\varphi_{\alpha}(\vec{r}_{\alpha i})
\nonumber \\
\times 
\prod_{\alpha = 1}^{\chi} 
{\cal{A}}(\phi_1(\vec{r}_{\alpha 1}),\cdots,
\phi_{N_{\alpha}}(\vec{r}_{\alpha N_{\alpha}})),
\end{eqnarray}
where ${\cal{A}}$ denotes the anti-symmetrizer and 
$f_{\alpha \beta}$, $\varphi_{\alpha}$ and 
$\phi_i$ denote two- or one-body functions whose functional form is
discussed in detail
in the following.

The $f_{\alpha \beta}$ denote two-body correlation factors.
For small $r$, $f_{\alpha \beta}$ coincides with the 
zero-energy scattering wave function for two particles
interacting through $V_{\alpha \beta}$. Beyond some
matching value $R_{\alpha \beta, m}$, which is treated as a variational
parameter, $f_{\alpha \beta}$ is taken
to be $c_{\alpha \beta, 1}+c_{\alpha \beta,2} \exp(- c_{\alpha \beta} r)$, 
where $c_{\alpha \beta}$ denotes a variational
parameter and $c_{\alpha \beta,1}$ and $c_{\alpha \beta, 2}$ 
are determined by the condition that 
$f_{\alpha \beta}$ and its derivative are continuous at $r=R_{\alpha \beta,m}$.
The $R_{\alpha \beta, m}$ and $c_{\alpha \beta}$ are optimized under the
constraint that $f_{\alpha \beta} \ge 0$ for all $r$. 

The $\varphi_{\alpha}$ denote Gaussian orbitals with widths $b_{\alpha}$,
$\varphi_{\alpha}(\vec{r}) = \exp(-r^2/(2 b_{\alpha}^2))$.
The parameters $b_{\alpha}$ control the size of the Fermi cloud,
and are optimized variationally.
In the non-interacting case, $b_{\alpha}=a_{ho}$. For 
$a_{\alpha \beta}<0$, the system becomes more
compact due to the attractive interactions, resulting in a 
smaller optimal value of $b_{\alpha}$.
If all $m_{\alpha}$ and all $b_{\alpha}$ are the same, i.e.,
$m_{\alpha}=m$ and $b_{\alpha}=b$, then
the product 
$\prod_{\alpha=1}^\chi\prod_{i=1}^{N_\alpha}\varphi_\alpha(\vec{r}_{\alpha
i})$ reduces to $\exp(-N R^2/(2b^2))$ and
the variational parameter $b$ is proportional to the mean
hyperradius of the variational wave  function.

The $\phi_i(\vec{r})$ are given by 
$H_{n_x}(x) H_{n_y}(y) H_{n_z}(z)$, where  $(n_x,n_y,n_z)=(0,0,0)$
for $i=1$, $(n_x,n_y,n_z)=(1,0,0)$ for $i=2$,
$(n_x,n_y,n_z)=(0,1,0)$ for $i=3$ and so on, and the $H_n$ denote
Hermite 
polynomials of degree $n$.

The anti-symmetry and nodal surface
of $\psi_T$ are determined by the $\chi$ Slater determinants
[second line of Eq.~(\ref{eq_trial})],
which contain
the one-body functions $\phi_i$.  
The nodal surface of $\psi_T$ coincides with that of the non-interacting 
multi-component Fermi gas. 
For small $|a_{\alpha \beta}|$, $a_{\alpha \beta} < 0$,
this is expected to be a 
very good approximation. 
It has been 
shown that the
nodal surface used here also provides reasonably tight
bounds for the energies of trapped
two-component gases~\cite{blum07,stec07b,chan07}
all
the way to unitarity.
Whether this holds true for multi-component Fermi gases
with $\chi > 2$ should be addressed in more detail in follow-up work.
Note that the guiding function $\psi_T$ used throughout this
work does not build in any ``pairing physics'' from the outset
(see Sec.~\ref{sec_conclusion} for further discussion).
The product
$\phi_{\alpha}(\vec{r}) \varphi_i(\vec{r})$,
$i=1,\cdots,N_{\alpha}$, with $b_{\alpha}=a_{ho}$ coincides with
the $i$th (non-normalized)
single-particle harmonic oscillator wave function.
We choose to write $\psi_T$ in terms of the product
$\phi_{\alpha} \varphi_i$ instead of the harmonic oscillator functions
themselves, because this allows the widths 
$b_{\alpha}$ of the Gaussians to be varied without
changing the nodal surface of $\psi_T$.

In addition to the VMC method, we apply the FN-DMC method~\cite{hamm94,reyn82}. 
The FN-DMC method uses the variational wave function
$\psi_T$ as a so-called guiding function and determines the energy of a
state whose nodal surface is identical to that of $\psi_T$. 
It can be shown that the FN-DMC energy
provides an upper bound to the lowest eigenenergy of the eigenstate 
that has the same symmetry as $\psi_T$~\cite{reyn82}. 
If the nodal surface of $\psi_T$ coincides with that
of the true eigenfunction, then the FN-DMC method results, 
within the statistical uncertainty, in the 
exact eigenenergy. 
In general, the quality of the FN-DMC energy depends crucially
on the construction of the nodal surface  of $\psi_T$.
As in the
VMC method, expectation values 
calculated by the FN-DMC method have a statistical uncertainty,
which can be reduced by increasing the computational efforts.

\subsection{Monte Carlo results}
\label{sec_mcresults}
We first consider three- and four-component Fermi gases with
one atom per component
and equal masses and trapping frequencies. 
If all interspecies scattering lengths are equal,
these Fermi gases are described by the
same Hamiltonian as the corresponding Bose
gas
and 
the system should become unstable towards the formation 
of negative energy states, characterized by small interparticle distances,
when the inequality given in Eq.~(\ref{eq_critbose}) is fulfilled.
For $N=3$ and 4, this implies critical scattering lengths
of $a_s \approx -0.29 a_{ho}$ and 
$-0.19 a_{ho}$, respectively.

To investigate how this instability or
collapse arises within the many-body Monte Carlo
framework, we perform DMC calculations for the three- and four-particle
systems interacting through
a square
well potential with $R_0=0.01a_{ho}$
(the subscripts $\alpha$
and  $\beta$ of $R_{\alpha \beta,0}$ have been dropped for
notational convenience).
The depth is chosen so that $a_s$ takes the desired value and so
that the potential supports no
two-body bound states. 
For small $|a_s|$, 
the DMC method results in 
energies that agree very well with the
mean-field Gross-Pitaevskii (GP) energy for both $N=3$ and 4.
As $a_s$ becomes more negative
(of the order of $-0.06a_{ho}$~\cite{footnotetimestep}), 
the DMC walkers sample at first
exclusively
positive energy configurations;
after a certain propagation time
some walkers sample configurations with negative energy
that correspond to tightly-bound states. 
The existence of tightly-bound three-body states, which depend on the
details of the underlying two-body potential, for 
scattering lengths $a_s$ that are less negative 
than the critical scattering lengths determined at the mean-field
level or within the hyperspherical framework
was already pointed out in Sec.~\ref{sec_hyper_3}. Our calculations here
show that the DMC algorithm ``sees'' 
these three-body bound states (as well as four-body bound states)
if the simulation time is
sufficiently long and $|a_s|$ is sufficiently large.
If $|a_s|$ is not too large, the DMC algorithm does---
if the initial walker configurations correspond to gas-like
states--- not ``know'' about
the tightly-bound states with negative energy.

\begin{figure}
\vspace*{.075in} 
\includegraphics[angle=270,width=80mm]{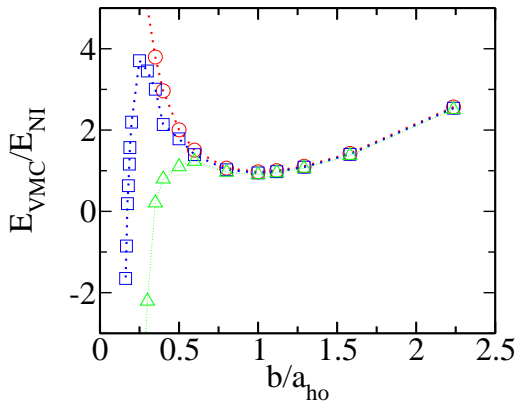}
\caption{
(Color online)
Circles, squares and triangles show the 
variational energy $E_{VMC}$
for $N=3$ atoms
interacting through a square well potential (range $R_0=0.01a_{ho}$)
with $a_s=-0.05a_{ho}$,
$-0.1a_{ho}$ and $-0.15a_{ho}$, respectively, as a function of 
the variational parameter $b$ [see the discussion 
following Eq.~(\protect\ref{eq_trial})].
$E_{VMC}$ is scaled by the energy $E_{NI}$ of the non-interacting
system, $E_{NI}=4.5 \hbar \omega$.
Dotted lines connect data points for ease of viewing.
}
\label{fig3a}
\end{figure}

To gain further insight, we treat the $N=3$ system by the VMC method. 
For a given $a_{s}$, 
we fix the variational parameters $R_{m}$
and $c$ (their exact values are not
important for the discussion that follows)
and vary $b$ 
(we drop the subscripts $\alpha$ and  $\beta$ 
of
$R_{\alpha \beta,m}$ and $c_{\alpha \beta}$, 
and the subscript $\alpha$ 
of
$b_{\alpha}$). 
Circles, squares and triangles
in Fig.~\ref{fig3a} show the variational energy $E_{VMC}$
for $N=3$
as a function of 
$b$ for $a_{s}=-0.05 a_{ho}$, $-0.1a_{ho}$ 
and $-0.15 a_{ho}$, respectively.
For $a_{s}=-0.05a_{ho}$,
$E_{VMC}$ increases if the Gaussian width $b$ is larger than
about $a_{ho}$ (the system
expands compared to its optimal size, thereby reducing the 
attraction felt between the atoms)
and if $b$ is smaller than about $a_{ho}$
(the system shrinks 
compared to its optimal size, thereby increasing the kinetic energy). 
For $a_{s}=-0.1 a_{ho}$ and $-0.15a_{ho}$
(squares and triangles in Fig.~\ref{fig3a}), 
the variational energy assumes a local minimum
at $b \approx a_{ho}$, shows a local maximum at 
$b \approx 0.25 a_{ho}$ for
$a_s=-0.1a_{ho}$ and at $b \approx 0.5a_{ho}$ for 
$a_s=-0.15a_{ho}$, and becomes negative for
yet smaller $b$. 
We refer to 
the local maximum 
as an ``energy barrier'' that separates the local minimum at
$b \approx a_{ho}$ from a global minimum 
that exists at smaller $b$ values.
Finally, for 
more negative scattering lengths
(not shown in Fig.~\ref{fig3a}) 
no energy barrier exists
for the variational
wave 
functions
considered.

The energy barrier discussed here for small 
$s$-wave interacting Bose gases
also exists in three-dimensional dipolar Bose gases~\cite{rone06} 
and one-dimensional
Bose gases~\cite{astr04}
(the second part of the Appendix in Ref.~\cite{rone06}
provides a detailed discussion). 
As in those earlier studies, we interpret the existence 
of the energy barrier as an indication that the Hilbert space 
is divided into two nearly orthogonal spaces if $a_s$
is quite a bit less negative than the critical scattering length
predicted at the mean-field level: Gas-like states
live in one region of the Hilbert space, and cluster-like 
bound states in the other.
While this reasoning leads to an intuitive
understanding of the stability and decay of Bose gases with
negative scattering length, the
question arises why the energy barrier disappears for scattering lengths
$a_s$ that are notably
less negative than the critical scattering length predicted by the GP equation.
We believe that this is due to the existence of
tightly-bound states with negative energy.
If $|a_s|$ is sufficiently large,
the overlap of the variational wave function 
with eigen functions of
cluster-like bound states increases with decreasing $b$.
For non-vanishing overlap, $E_{VMC}$ 
provides a rather poor upper bound to the true ground state
energy of the system and not
an upper bound to the energetically lowest-lying gas-like state.
This implies that our VMC calculations do not allow
for a reliable determination of the critical scattering length. Instead,
they 
indicate that the existence of the energy barrier gives rise
to the stability
of the gas-like state and that 
the disappearing of the energy barrier qualitatively
explains how the instability of Bose gases 
and multi-component Fermi gases with one atom per component
arises when $a_s$ becomes too negative.

We note that the picture developed here based on our VMC and DMC calculations
is closely related to the analysis 
based on the
hyperspherical formulation presented in Sec.~\ref{sec_hyper}
and in Refs.~\cite{bohn98,hart00,blum02b,ritt07}.
In the VMC calculations, the role
of the hyperradius $R$ is played by the Gaussian width $b$
[see discussion following Eq.~(\ref{eq_trial})]. 
Furthermore,
variational mean-field
calculations have been interpreted in much the same 
way~\cite{pere97,stoo97}.

We now investigate the behaviors of multi-component 
Fermi gases with more than one atom per component for which all 
interspecies scattering lengths $a_{\alpha \beta}$ with $\alpha \ne \beta$
are resonant; as before, we set $a_{\alpha \beta} =a_s$.
We first solve 
the many-body
Schr\"odinger equation for a square well potential with range
$R_{\alpha \beta,0} = 0.01 a_{ho}$ by the FN-DMC method.
\begin{figure}
\vspace*{.075in} 
\includegraphics[angle=270,width=80mm]{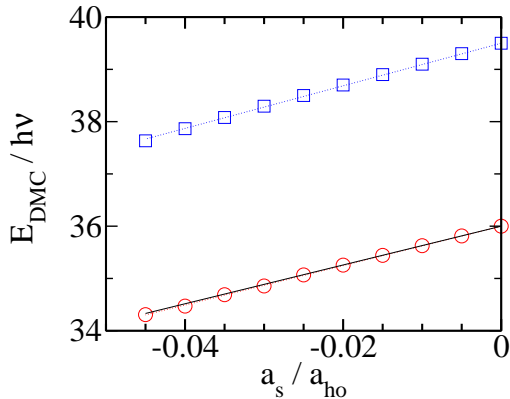}
\caption{
(Color online)
Circles and squares show the 
FN-DMC energy $E_{DMC}$
for a four-component Fermi gas with $N=16$ and $17$ atoms,
respectively, interacting through a square well potential with range
$R_0=0.01 a_{ho}$ as a function of the interspecies scattering length
$a_{s}$ (all interspecies scattering lengths are equal).
While the $N=16$ system has four atoms per component,
the $N=17$ system has one component with five atoms
and three components with four atoms.
Dotted lines show a linear fit to the FN-DMC energies: 
The slope is $37.7 \hbar \omega / a_{ho}$ for $N=16$ and
$40.7 \hbar \omega / a_{ho}$ for $N=17$.
For comparison, the solid line shows the energy for 
$N=16$ 
calculated within first order perturbation theory, 
$E=36\hbar\omega+ca_{s}/a_{ho}$
with 
$c \approx 37.1 \hbar \omega$.
}
\label{fig5}
\end{figure}
Circles in Fig.~\ref{fig5} show the FN-DMC energy for a four-component
Fermi gas with four atoms per component (this corresponds to a closed
shell) as a function of the $s$-wave scattering length $a_{s}$.
The energy decreases approximately linearly with decreasing 
$a_{s}$, suggesting that this weakly-interacting system
can be described to a good approximation
perturbatively. 
Assuming zero-range interactions
with scattering lengths $a_s$ and treating the system
within first order perturbation theory, we find
$E\approx 36 \hbar \omega+c a_{s}/a_{ho}$, where
$c=93  \hbar \omega / \sqrt{2 \pi}$,
for $N=16$.
This perturbatively calculated energy (solid line in Fig.~\ref{fig5})
describes the FN-DMC energies (circles) very well. 
By additionally treating the $N=17$ fermion system 
with $N_1=5$ and $N_2=N_3=N_4=4$ (squares
in Fig.~\ref{fig5}), we find that the chemical potential decreases,
just as the energy, approximately linearly with
increasing $|a_{s}|$.
For scattering lengths more negative than about 
$-0.045a_{ho}$~\cite{footnotetimestep}, 
the DMC walkers sample---
just as in the case of the small Bose systems--- negative 
energy configurations. 
We find that the three-component Fermi gas
with $N=12$ behaves similarly. 
At the FN-DMC level, the instability of multi-component Fermi gases
is accompanied by the existence of cluster-like states
with negative energy.
To investigate whether these cluster-like states ``live'' for sufficiently
small $|a_{s}|$, as in the case
of Bose gases, in a Hilbert space that is approximately
orthogonal to the Hilbert space where the gas-like states live,
we perform a series of VMC calculations.

\begin{figure}
\vspace*{.075in} 
\includegraphics[angle=270,width=80mm]{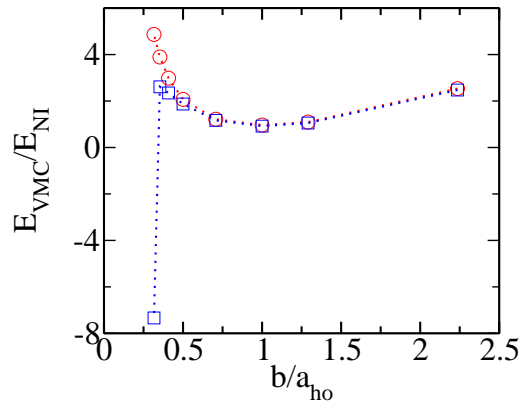}
\caption{
(Color online)
Circles and squares show the 
variational energy $E_{VMC}$
for a three-component Fermi gas with $N=12$ atoms
interacting through a square well potential (range $R_0=0.01a_{ho}$)
with $a_{s}=-0.05a_{ho}$ and 
$-0.1a_{ho}$ (all interspecies scattering lengths
are equal), respectively, as a function of 
the variational parameter $b$.
$E_{VMC}$ is scaled by the energy $E_{NI}$ of the non-interacting
system, $E_{NI}=27 \hbar \omega$.
Dotted lines connect data points for ease of viewing.
}
\label{fig3}
\end{figure}

\begin{figure}
\vspace*{.075in} 
\includegraphics[angle=270,width=80mm]{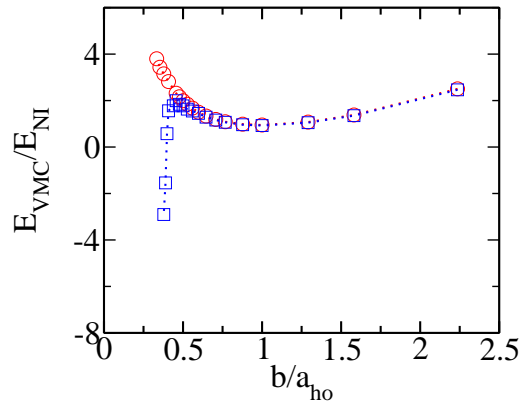}
\caption{
(Color online)
Circles and squares show the 
variational energy $E_{VMC}$
for a four-component Fermi gas with $N=16$ atoms
interacting through a square well potential (range $R_0=0.01a_{ho}$)
with $a_{s}=-0.05a_{ho}$ and 
$-0.07a_{ho}$,
respectively, as a function of 
the variational parameter $b$.
$E_{VMC}$ is scaled by the energy $E_{NI}$ of the non-interacting
system, $E_{NI}=36 \hbar \omega$.
Dotted lines connect data points for ease of viewing.
}
\label{fig4}
\end{figure}

Circles and squares in Fig.~\ref{fig3} show the variational energy $E_{VMC}$
for the three-component Fermi system with $N=12$
as a function of 
$b$ 
for $a_{s}=-0.05 a_{ho}$
and $-0.1 a_{ho}$, respectively
(as before, $b=b_{\alpha}$ and $a_s = a_{\alpha \beta}$ where 
$\alpha \ne \beta$).
The lowest variational energy for $a_{s}=-0.05 a_{ho}$
is obtained 
for $b \approx a_{ho}$, $E_{VMC}= 26.09(4)\hbar \omega$. 
The FN-DMC energy for this interspecies scattering 
length, $E_{DMC}=26.08(4) \hbar \omega$, 
agrees with $E_{VMC}$ within error bars, 
indicating that our variational wave function---
assuming the nodal surface is adequate---
captures the key physics of the system.
For comparison,
circles and squares in Fig.~\ref{fig4} show the variational energy $E_{VMC}$
for the four-component Fermi system with $N=16$
as a function of 
$b$ for $a_{s}=-0.05 a_{ho}$
and $-0.07 a_{ho}$, respectively.
A comparison of Figs.~\ref{fig3} and \ref{fig4} shows that the overall
behavior of the three- and four-component systems 
is very similar, and resembles that
discussed above for Bose gases:
For small $|a_s|$,
$E_{VMC}$ increases if the Gaussian width $b$ is larger than
about $a_{ho}$ 
and if $b$ is smaller than about $a_{ho}$.
For somewhat more negative interspecies scattering lengths $a_s$,
$E_{VMC}$
shows a local minimum
at $b \approx a_{ho}$ and a local maximum at 
$b \approx 0.3 a_{ho}$, and $E_{VMC}$
becomes negative for
yet smaller $b$. 
For more negative $a_s$ (not shown in Figs.~\ref{fig3} and \ref{fig4}),
the energy barrier disappears.
We note that the
variational energies for $b$ values
of the order of $0.3a_{ho}$ to $0.6a_{ho}$ and 
$a_{s} \lesssim -0.05a_{ho}$
show large fluctuations.
For these systems, configurations with rather different 
``geometries'', and hence with rather different
potential and kinetic energies, 
are being sampled, leading to non-Gaussian distributed 
energy expectation values. This behavior is a consequence of the 
fact that the variational wave function does not provide a good
description of the true eigenfunction.

Our MC results for multi-component Fermi gases with all
resonant interactions 
and $\chi>2$
support
the physical picture developed in Sec.~\ref{sec_hyper} 
within the hyperspherical 
framework about how
the instability or ``implosion'' of the system arises.
Furthermore,
the VMC and FN-DMC results suggest that the stability
and decay of multi-component Fermi gases can be attributed to the
same mechanisms as the stability and
decay of Bose gases. 
At the VMC level,
the energy barrier arises on length scales that are
comparable to $|a_{s}|$ but 
much larger than the range 
of the two-body potential. This suggests that
the onset of the instability is determined by the value
of the scattering length as opposed to the details of the underlying
two-body potential,
assuming the range of the under-lying two-body potential
is sufficiently small (see also the discussion in Sec.~\ref{sec_hyper_3}).

\begin{figure}
\vspace*{.075in} 
\includegraphics[angle=270,width=80mm]{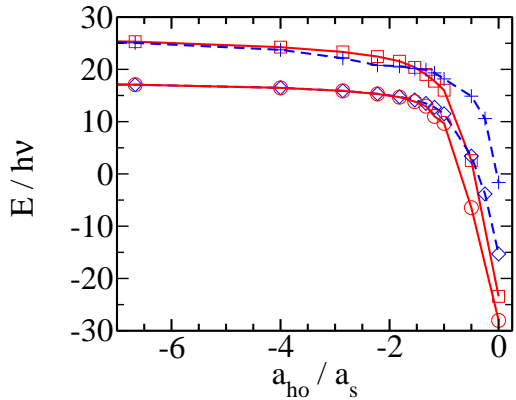}
\caption{
(Color online) 
Energies for a three-component Fermi system as a function
of $a_{ho}/a_s$ for two resonant interactions.
Circles and diamonds show the energy of the
three-particle system, multiplied by a factor of four,
interacting through
a square well potential with range $0.01a_{ho}$ and
a Gaussian potential with range $0.007a_{ho}$, respectively;
the energies  for the square well potential
are calculated by the FN-DMC approach and those for the
Gaussian potential by the CG 
approach.
Squares and pluses show the FN-DMC energies for 
$N=12$ for
a square well potential with range $0.01a_{ho}$
and a Gaussian potential with range $0.007a_{ho}$, respectively.
Solid and dashed lines connect data points for ease of viewing.
}
\label{fig_twores}
\end{figure}

We now discuss the behavior of three-component Fermi gases
with equal masses and equal trapping frequencies
in which two interspecies scattering lengths,
denoted in the following by $a_s$,
are resonant
and the third
interspecies scattering length is zero.
As before, the depths of the two-body potentials
are adjusted so that the
scattering lengths take the desired value and so that
the two-body potentials support no two-body bound state
(for $a_{\alpha \beta}=0$, the depth is set to zero).

Motivated by the discussion presented in Sec.~\ref{sec_hyper_3}
for the three-particle system we first consider 
two-body potentials with very small range.
The trimer system interacting through a square well potential with
$R_0=0.001a_{ho}$ supports a state with negative energy for scattering lengths
that are much less negative than the critical scattering length of
$a_c \approx -1a_{ho}$ predicted for $N=3$
(see Sec.~\ref{sec_hyper_3}). Thus, we expect the stability
behavior of
the three-component many-fermion 
system with two resonant interactions 
interacting through square well potentials
with range $R_0=0.001 a_{ho}$ to be similar
to that of the all-resonant system discussed above.
Indeed, we find that the FN-DMC calculations for the $N=12$ system
appear stable for small $|a_s|$, including a 
region of scattering lengths where the three-particle
system supports negative energy states, but show instabilities for larger
$|a_s|$ ($|a_s| < |a_c|$).
The VMC energy is minimal for $b \approx a_{ho}$ and increases 
for smaller and larger $b$ values. Somewhat surprisingly, 
our VMC calculations do not indicate the existence of an energy
minimum for small $R$ values when $a_s$ is
close to $a_c$. This is different from 
the all-resonant systems and can possibly be attributed
to the variational wave functions employed (for 
very
small $R_0$,
a more flexible functional form may be needed to obtain an energy
minimum at small $R$).
We find similar results for the four-component system with three
resonant interactions for which $a_{1\beta} = a_s$ and $a_{\alpha \beta}=0$ 
otherwise (we did not perform Monte Carlo calculations for the second
non-trivial configuration with $\chi-1$ resonant interactions, 
for which $a_s=a_{12}=a_{23}=a_{34}$ and $a_{\alpha \beta}=0$ otherwise).
Our FN-DMC results for multi-component Fermi gases
with $\chi-1$ resonant interactions 
($\chi > 2$)
and very small $R_0$ suggest that these
systems
become 
unstable
for a certain critical negative scattering length. 
In analogy to the Bose system, this seems to suggest
that multi-component Fermi gases with very small $R_0$ 
are stable for scattering lengths
less negative than $a_c$ and become mechanically unstable
for scattering lengths more negative than $a_c$.

Next, we consider three-component Fermi gases interacting through two-body
potentials with a somewhat larger range with $a_s=a_{12}=a_{23}$ and 
$a_{31}=0$.
Figure~\ref{fig_twores} shows the energy for the three-particle
system, multiplied by a factor of four, as a function
of $a_{ho}/a_s$ for the square well potential with $R_{0}=0.01a_{ho}$
(circles) and the Gaussian
potential with $R_0=0.007a_{ho}$ (diamonds). 
The $N=3$ energies for the Gaussian potential are calculated by the
CG approach~\cite{stec07c,stec07,stec07b,blum07}  
(this approach is free of any assumptions, and its
accuracy can be improved systematically) and those for
the square well potential by the FN-DMC approach.
For small $|a_s|$, the energies for the two 
different interaction potentials agree quite well
but for large $|a_s|$ deviations
are visible. Thus, the system's behavior is,
although qualitatively similar, to some extent non-universal.
The three-body energy becomes negative for
$a_s \approx -1.5a_{ho}$ and $-3a_{ho}$ 
for the square well and Gaussian potentials, respectively.

The energies for the three-component Fermi system with four atoms
per component, i.e., for $N=12$, are shown in Fig.~\ref{fig_twores} 
by squares and pluses for the
square well and the Gaussian potential, respectively.
As in the three-particle case, the energies for these two different
potentials agree well for small $|a_s|$ but behave somewhat differently
for large $|a_s|$. 
For both potentials, the $a_s$ value at which negative energy states
appear is similar to that for the corresponding
three-body system.
An analysis of the structural properties suggests that the $N=12$ system
is made up of weakly-bound three-body clusters.
Figure~\ref{fig_twores} shows 
that the energy for $N=12$ is larger than four times
the energy of the corresponding trimer system,
suggesting that the many-body system, which appears to be made
up of weakly-bound trimers, may be stable.
Our FN-DMC calculations show no evidence for the formation of negative
energy states consisting of clusters that contain
more than three atoms.
At this point it is not clear whether this is a consequence
of the particular guiding function employed or whether 
many-body clusters consisting of four, five or six atoms
with negative energy do indeed not
exist.

\section{Summary and conclusion}
\label{sec_conclusion}

\begin{table}
\caption{\label{tab2} 
Summary of our stability analysis:
In the large $|a_s|$ limit,  trapped $\chi$-component
Fermi gases
are predicted to be either 
stable  (``S'') or unstable (``U''), or to form
a (possibly stable) gas of trimers (``S(?)GT'').
The predictions are based on the hyperspherical treatment,
which employs the bare and the 
density-dependent
pseudo-potential,
and a many-body Monte Carlo framework.
$^{(1)}$Only systems 
with $a_s=a_{12}=a_{13}=a_{14}$ and $a_{\alpha \beta}=0$ otherwise
with very small two-body range were
investigated.}
\begin{ruledtabular}
\begin{tabular}{l|ccc}
$\chi$  & hyperspherical & hyperspherical & FN-DMC/  \\
 & (``bare'' PP) & (den. dep. PP) & VMC  \\ \hline
$2$ & U & S & S \\
$3$ (all res.) & U & U & U \\
$3$ (two res.) & U & S & U or S(?)GT \\
$4$ (all res.) & U & U & U \\
$4$ (three res.) & U & S & U$^{(1)}$ 
\end{tabular}
\end{ruledtabular}
\end{table}

This paper investigates the stability of trapped $\chi$-component 
Fermi gases interacting through short-range two-body
potentials with negative scattering lengths. Table~\ref{tab2}
summarizes the results of our stability analysis.

One of our main findings is that trapped multi-component Fermi gases
with more than two components, in which all 
scattering lengths, masses and trapping frequencies 
are equal, become unstable for a certain critical negative scattering length.
This instability, a type of 
Ferminova, is similar to the instability of trapped Bose gases
with negative scattering length, sometimes referred to
as bosenova~\cite{donl01}.
The instability predicted here for
multi-component Fermi systems, $\chi>2$, 
emerges within two different theoretical frameworks:
A hyperspherical many-body treatment that employs 
either the bare or a 
density-dependent zero-range interaction~\cite{ritt07,stec07},
and a many-body Monte Carlo treatment, which includes both
VMC and FN-DMC calculation and employs finite-range
two-body interactions with a range $R_0$ chosen so that the 
first 
three-particle state with negative energy
appears for $a_s$ much less negative than $a_c$.

Both the hyperspherical and the Monte Carlo 
frameworks parameterize the nodal surface of the
anti-symmetric many-body wave function by the ideal gas nodal surface.
In the limit of vanishing confinement and periodic
boundary conditions, this corresponds to a 
wave function that describes a normal state.
This paper does not investigate wave functions that
include pairing physics 
or the possibility of phase separation 
from the outset. For three-component
systems, for instance,
a BCS mean-field framework predicts the existence of a
pairing mechanism in which
the atoms of species one and two pair (forming a superfluid)
while the atoms of species three are in 
a normal state~\cite{chan06,he06,paan07,zhai07}.
This pairing mechanism has also been investigated within
other theoretical frameworks~\cite{hone04,hone04b,beda06,cher07}.
Although the parameter combinations considered in the above cited references
is different from those considered in the present paper, we believe
that future Monte Carlo studies of multi-component 
Fermi gases with more than two components should investigate the implications
of a many-body wave function that has pairing physics
build in from the outset. In particular,
it would be interesting to address the question whether a state 
that contains pairing physics would be stable against the collapse 
investigated 
in the present study. If such a 
stabilizing mechanism existed this would 
have important 
consequences for on-going cold atom experiments. 

To observe the predicted collapse of multi-component 
Fermi gases with all-resonant  interactions,
we imagine the following experiment.
First, a stable $\chi$-component Fermi gas 
($\chi > 2$) with vanishing interspecies scattering lengths
is prepared.
We then imagine that all interspecies scattering lengths be 
suddenly tuned to the same large negative value. The time scale for this
ramp should be fast compared to the time scale at which phase separation
might occur and also fast compared to the time scale at which losses due
to spin-flip collisions
might occur.
The attraction between particles would then produce an implosion and collapse
of the system.
Following the implosion, we
anticipate that recombination into molecules and clusters will immediately
follow once the system reaches high density, and the resulting energy
release should detonate a ferminova akin to the bosenova that has been
observed experimentally and discussed 
theoretically~\cite{donl01,rupr95,pere97,brad97,sack99,gama01}.
The rich dynamics of soliton formation, akin
to that observed for bosonic atoms in Ref.~\cite{corn06} 
could also conceivably ensue.

A second main finding of our work concerns three-component 
Fermi systems with two resonant and one non-resonant
interspecies scattering lengths. The behavior of these systems
depends strongly on the range $R_0$ of the underlying two-body potential.
For very small $R_0$, an instability similar to that for the three-component
system with all-resonant interactions 
appears to arise,
although at much more negative
$a_s$. For larger $R_0$, the many-body  system
may be made up of weakly-bound trimers and 
may be stable all the way up to unitarity.
We note that the importance of the range parameter, or a three-body
parameter, was already pointed out in earlier 
work~\cite{beda06,chan06,sedr06}. Future studies should investigate
in more detail whether a Fermi system consisting of weakly-
or strongly-bound trimers could be realized
experimentally.

Finally, we note that multi-component Fermi gases have recently
also been investigated in one-dimensional space~\cite{lech05,guan07,liu07}.
Assuming zero-range interactions, many properties of the multi-component
system can be obtained analytically through the exact Bethe ansatz.
For negative coupling constant, the ground state of $\chi$-component
Fermi gases contains clusters
that consist of $\chi$-atoms (one of each species). 
It is suggested that one-dimensional three-component Fermi gases
undergo a phase transition that is analogous to the quark color
superconductor state~\cite{liu07}.
In the future, it would be interesting to investigate whether
such a transition also exists in three-dimensional
three-component Fermi gases.

Clearly, more work is needed to better understand the behavior of
three-dimensional multi-component Fermi gases, including the
implications of different two-body ranges and different 
nodal surfaces of the guiding function employed in the Monte Carlo
study.
This work investigated the behavior of multi-component Fermi 
gases for different scattering lengths and two-body ranges. 
In cases where three-body states with negative energy exist, 
it will be important to determine if the range of the two-body parameter is
indeed the relavant parameter; it may be that the
trimer binding energy is a more natural quantity.

We gratefully
acknowledge
support by the NSF
and discussions with J. D'Incao.


\begin{thebibliography}{91}
\bibitem[{\citenamefont{{In fact, these systems are metastable; however,
  throughout this paper we call two-component Fermi gases--- owing to their
  long lifetime--- stable.}}(?)}]{footnotestable}
\bibinfo{author}{\bibnamefont{{In fact, these systems are metastable; however,
  throughout this paper we call two-component Fermi gases--- owing to their
  long lifetime--- stable.}}}
\bibitem
{footnoterenorm}
\bibinfo{author}{\bibnamefont{The hyperspherical treatment presented in
  Ref.~\cite{ritt07} uses the value $\beta=-0.49$~\cite{stec07} instead of
  $\beta=-0.58$. As a consequence, Ref.~\cite{ritt07},
  predicts that all-resonant three-component systems are
  stable all the way to unitarity while the present analysis predicts a
  collapse due to the more negative $\beta$ value.}}
\end{thebibliography}
\end{document}